\documentclass[12pt]{article}
\usepackage{epsfig}
\usepackage{amsfonts}
\usepackage{amsopn}
\usepackage{amsmath}
\usepackage{verbatim,amsthm}
\title
{\vskip -50 pt
\begin{flushright}
\normalsize\rm NORDITA-2016-23
\end{flushright}
\vskip 20 pt
 Phenomenological Lagrangians, gauge models and branes
}

\author{
 A. A. Zheltukhin 
  \thanks{E-mail: aaz@physto.se} 
  \\
Kharkov Institute of Physics and Technology, \\
1, Akademicheskaya St., Kharkov, 61108, Ukraine \\  
NORDITA, Royal Institute of Technology and Stockholm University,\\
Roslagstullsbacken 23, SE-106 91 Stockholm, Sweden 
}

\date{}
\begin{document}

\maketitle

\begin{abstract}
Phenomenological Lagrangians for physical systems with 
spontaneously broken symmetries are reformulated in terms 
of gauge field theory.
Description of the Dirac $p$-branes in terms of the Yang-Mills-Cartan 
gauge multiplets interacting with gravity, is proved 
to be equivalent to their description
as a closed dynamical system with the symmetry $ISO(1,D-1)$ 
spontaneously broken to $ISO(1,p)\times SO(D-p-1)$.

\end{abstract}

The geometric approach introduced for the description 
of string [1-5] \nocite{Car, RL, Om, BNCh, BN} 
considers its worldsheet as a surface embedded into 4-dim. 
Minkowski space $\mathbf{R}^{1,3}$. 
The gauge reformulation  \cite{Zgau24}, \cite{Zbran} 
of the geometric approach represents the action of strings 
and $p$-branes in terms of the interacting Yang-Mills-Cartan 
multiplets and gravitational field localized on a $(p+1)$-dim.  
world hypersurface $\Sigma_{p+1}$ swept in $\mathbf{R}^{1,D-1}$. 
 Using the Cartan formalism of moving frames  
 \cite{Car} and its physical development 
 [8-10] \nocite{Vol1, VZufg, Zantis}
 we interprete the above gauge description in terms of the 
 Nambu-Goldstone (N-G)  fields of the spontaneously broken Poincare symmetry 
 $ISO(1,D-1)$ studied in [11-16] \nocite{BCGM, GKW, CLNVX, GM, AKo, GKP}.
Then string and  p-brane emerge as the general solutions 
of the Euler-Lagrange EOM selected by the Maurer-Cartan 
eqs.  that play the role of the Cauchy-Kovalevskaya initial data.
We show that the generalized eqs. of the {\it Gauss Theorema Egregium} 
are the dynamical eqs. for the string and brane metrics. 
For the string embedded into  $\mathbf{R}^{1,2}$ these eqs.  
represent the gravity described by the geometry of 2-dim. Einstein space.

\bigskip 

1. Consider a global semisimple group of symmetry $G$ 
 spontaneously broken to $H$
\begin{eqnarray}\label{comr}
[Y_{\alpha}, Y_{\beta}]= ic^{\gamma}{}_{\alpha\beta}Y_{\gamma}, \ \
[X_{i}, Y_{\alpha}]= ic^{k}{}_{i\alpha}X_{k}, \ \
[X_{i}, X_{k}]= ic^{\alpha}{}_{ik}Y_{\alpha} + ic^{l}{}_{ik}X_{l}.
\end{eqnarray}
and use the following factorized representation of its group elements
 \cite{Vol1}, [18-21] \nocite{Wei, Schw, CWZ, CCWZ} 
\begin{eqnarray}\label{fctr} 
G(a,b)= K(a)H(b),
\end{eqnarray}
where $a$ and $b$ parametrize the group space of $G$.
The left multiplication 
 \begin{eqnarray}\label{actn}
 gG=G'  \ \ \    \rightarrow  \ \ \ \ gK(a)H(b)=K(a')H(b')  
 \end{eqnarray}
 yields the nonlinear transformation of the group parameters 
\begin{eqnarray}\label{trf} 
a'=a'(a,g),  \ \ \ \ b'=b'(b,a,g)
\end{eqnarray}
which  preserves the differential form $G^{-1}dG$ 
and its components 
\begin{eqnarray}\label{lshi}
  G'^{-1}dG'=G^{-1}dG =i\omega_{G}^{i}(a,b,da)X_{i} 
  + i\theta_{G}^{\alpha}(a,b,da,db)Y_{\alpha}.
\end{eqnarray}
The parameters $a$ may be mapped into the components of the N-G 
field $\pi(x)$ with the same nonlinear transformation law 
$\pi'(x)= a'(\pi(x),g)$ for the construction of $G$-invariant nonlinear 
phenomenological Lagrangians for $\pi(x)$. 

An alternative approach proposed for the chiral sigma models
in \cite{FST}, and generalized to strings in \cite{Zgau24}, 
is based on consideration of $G$ as 
 completely spontaneously broken symmetry when 
all group parameters $(a,b)$ are treated
 as N-G modes. This extension is accompanied with a {\it compensating}
 extension of the {\it left global} symmetry of Lagrangian
  by a new {\it right gauge} symmetry $H_{R}$.
 The latter is realized by the {\it right} multiplications $G' = Gh$ 
 with $h \in H_{R}$.
In view of this gauge invariance, the non-physical $b$-modes 
associated with the parameters 
of the vacuum subgroup $H$ are exluded by a gauge fixing. 
The transformation rules for $G$-invariant Cartan forms (\ref{lshi}) 
under the right transformations from  $H_{R}$ are 
\begin{eqnarray}\label{rgtr}
G' = G h \ \ \ \ \rightarrow \ \ \ \  G'^{-1}dG'= h^{-1}(G^{-1}dG)h + h^{-1}dh.
\end{eqnarray}
The substitution of expansion (\ref{lshi}) 
into (\ref{rgtr})  using (\ref{comr}) gives the following 
gauge transformations 
\begin{eqnarray}\label{fadtr}
\omega_{G}^{' i}X_{i} =\omega_{G}^{i}h^{-1}X_{i}h, \ \ \ 
\theta_{G}^{' \alpha}Y_{\alpha}= \theta_{G}^{\alpha}h^{-1}Y_{\alpha}h -ih^{-1}dh
\end{eqnarray}
for the left invariant one-forms $\omega_{G}^{i}$ and $\theta_{G}^{\alpha}$.
This yields the transformation law 
\begin{eqnarray}\label{inffad}
\delta\omega_{G}^{k} =-c^{k}{}_{i\beta}\epsilon^{\beta}\omega_{G}^{i},  \ \ \ \ \ \ \
\delta\theta_{G}^{\gamma}= d\epsilon^{\gamma} - c^{\gamma}{}_{\alpha\beta}\epsilon^{\beta}\theta_{G}^{\alpha}
\end{eqnarray}
for the infinitesimal 
transformations $h\approx 1 + i\epsilon^{\beta}Y_{\beta}$ 
with the field-dependent parameters $\epsilon^{\beta}$. 
This shows that these forms can be treated as 
 one-forms of the massless vector and gauge 
 multiplets of $H_{R}$ built from the N-G fields.
In the case of Dirac $p$-branes embedded
into the Minkowski 
space $\mathbf{R}^{1,D-1}$ its global 
Lorentz symmetry $SO(1,D-1)$ 
plays the role of the discussed internal 
symmetry $G$, i.e. $G=G_{Lor}\equiv SO(1,D-1)$.  
The Lorentz symmetry is spontaneously 
broken to $SO(1,p)\times SO(D-p-1)$ due to 
the  presence of brane. Thus, the rotational N-G fields 
can be described in terms of the left invariant 
one-forms (\ref{fadtr}). On the other hand, 
the rotational DOF of branes can be presented  
by the vectors $\mathbf{n}_{A}(\mathbf{x})$ of
 the Cartan moving frame in  $\mathbf{R}^{1,D-1}$, 
 where $\mathbf{x}=\{x^{m}\}, (m=0,1,...,D-1)$  are the 
 global Cartezian coordinates. It shows that 
  $\mathbf{n}_{A}(\mathbf{x})$ can encode the rotational N-G fields.

2.  The moving frame in $\mathbf{R}^{1,D-1}$ is formed by the orthonormal
 vectors  $\mathbf{n}_{A}(\mathbf{x})$  \cite{Car},
 \begin{eqnarray}
 \mathbf{n}_{A}(\mathbf{x})\mathbf{n}_{B}(\mathbf{x})=\eta_{AB}, 
 \ \ \ (A,B=0,1,..,D-1), \label{mfra}  \\
 d\mathbf{n}_{A}=-\omega_{A}{}^{B}(d)\mathbf{n}_{B},  \ \ \ \  
 d\mathbf{x}= \omega^{A}(d)\mathbf{n}_{A} \ \ \ \ \ \  
\nonumber
 \end{eqnarray}
with their  vertex at the point $\mathbf{x}$. 
So, the frame is defined as the pair  $(\mathbf{x}, \mathbf{n}_{A}(\mathbf{x}))$ 
called the moving $D$-hedron  \cite{GuZ}, \cite{BZ_hedr}. 
In view of the $\frac{1}{2}D(D+1)$ constraints  $D^2$ componets $n_{mA}$ 
of $\mathbf{n}_{A}$ represent the pseudoorthogonal 
matrices $\hat{n}=n_{mA}$ parametrized by  $\frac{1}{2}D(D-1)$ 
independent fields  $\pi^{\lambda}(\mathbf{x})$. 
So, $\mathbf{n}_{A}=\mathbf{n}_{A}(\pi^{\lambda})$, 
and $\pi^{\lambda}$ can be treated as N-G bosons of 
 the completely spontaneously broken $SO(1,D-1)$ symmetry
  realized by the left global multiplications 
 \begin{eqnarray}\label{Glob}
n'_{mA}=l_{m}{}^{k}n_{kA}, \ \ \ \ l_{m}{}^{p}l^{n}{}_{p}=\delta_{m}{}^{n}, 
\end{eqnarray}
where the matrices $\hat{l} \in SO(1,D-1)$. 
As an example, choose $SO_{R}(1,D-1)$ as such a gauge 
 group acting to the right index $A$ of $\mathbf{n}_{A}$  
\begin{eqnarray}\label{Lorg}
\mathbf{n'}_{A}=
- L_{A}{}^{B}(\pi^{\lambda})\mathbf{n}_{B}, \ \ \ \ L_{A}{}^{B}L^{C}{}_{B}
= \delta_{A}^{C}.
\end{eqnarray}
As a result, the matrix $n_{mA}$ becomes
double  covariant under the left 
and right shifts
\begin{eqnarray}\label{L-Rtr}
n'_{mA}=l_{m}{}^{k}n_{kB}L^{B}{}_{A}(\pi^{\lambda}).
\end{eqnarray} 
But, in this case $SO_{R}(1,D-1)$  removes all the rotational N-G fields.
It is instructive to understand  
this result in terms of the left form  (\ref{lshi}) invariant under the 
global Lorentz group  
\begin{eqnarray}\label{wform}
 \omega_{A}{}^{B}(d)= (\hat{n}^{-1}d\hat{n})_{A}{}^{B}\equiv\mathbf{n}_{A}d\mathbf{n}^{B}.
 \end{eqnarray}
 Infinitesimal transformtaions of $\omega_{A}{}^{B}(d)$ under the right 
 local rotations of $SO_{R}(1,D-1)$
 \begin{eqnarray}\label{Rtr}
 \delta\mathbf{n}_{A}=\mathbf{n}_{B}\epsilon^{B}{}_{A}  \ \ \  
 \end{eqnarray}  
turn out to be similar to the gauge transformations of a gauge field 
potential 
  \begin{eqnarray}\label{str}
  \delta\omega_{A}{}^{B}(d)=d\epsilon_{A}{}^{B} + [\omega(d), \epsilon]_{A}{}^{B}
 \end{eqnarray}
with the commutator $[\hat\omega, \hat\epsilon]$ in the  r.h.s. of the law. 
The covariant differential $D_{A}{}^{B}$ 
 \begin{eqnarray}\label{covde}
D_{A}{}^{B}=d\delta_{A}{}^{B} + \omega_{A}{}^{B}(d).
\end{eqnarray}
 for a vector field $\mathbf{V}= V^{A}\mathbf{n}_{A}$, 
 transforming as $\delta V_{A}= V_{B}\epsilon^{B}{}_{A}$, 
 has  the homogeneous law
 \begin{eqnarray}\label{covdif}
 \delta (DV)_{A}= (DV)_{B}\epsilon^{B}{}_{A}. 
\end{eqnarray}
The exterior product of differentials  (\ref{covde}) 
yields the gauge covariant 2-form $F_{A}{}^{B}(d)$ 
\begin{eqnarray}\label{covstr}
F_{A}{}^{B}:=D_{A}{}^{C}\wedge D_{C}{}^{B}= 
 (d\wedge\omega + \omega\wedge\omega)_{A}{}^{B}.
\end{eqnarray}
The description of the N-G fields in terms of the gauge 
field $\hat{\omega}$ requires the integrability of 
Eqs. (\ref{wform}) that 
can be rewritten in the form of the PDEs 
\begin{eqnarray}\label{wform'}
d\mathbf{n}_{A}=-\omega_{A}{}^{B}(d)\mathbf{n}_{B}.
\end{eqnarray}
The integrability conditions for the system (\ref{wform'}) have the form
\begin{eqnarray}\label{integr}
d\wedge\omega_{A}{}^{B} +\omega_{A}{}^{C}\wedge\omega_{C}{}^{B}=0  
\ \ \ \rightarrow \ \ \ F_{A}{}^{B}=0.
\end{eqnarray}
So, we see that in the  absence of the N-G fields the 
corresponding  potential $\hat{\omega}$ must be 
a pure gauge form. This illustrates the equality of 
the DOF 
carried by the N-G fields and the gauge field $\omega_{A}{}^{B}$. 
When the Lorentz symmetry is partially broken to its 
 vacuum subgroup $H \in SO(1,D-1)$, the right gauge 
 subgroup $H_{R} \in SO(1,D-1)_{R}$ must be used 
instead of  $SO(1,D-1)_{R}$.
 Then $H_{R}$ will remove only the N-G fields 
 corresponding to  the generators of the subgroup $H$. 
That explains how one can change 
the standard description of the N-G fields by the transition to  
the vector and gauge multiplets.
 
 For $p$-brane with the world vector
 $\mathbf{x(\xi)}=\{x^{m}(\xi)\}, (\xi^{\mu}=(\tau,\sigma^r), \, r=1,2,..,p)$ 
of  its world hypersurface $\Sigma_{p+1}$,
 the vacuum subgroup $H = SO(1,p)\times SO(D-p-1)$. 
That forces to choose $H_{R} = (SO(1,p)\times SO(D-p-1))_{R}$ as 
the right gauge group. The latter is composed from 
 the tangent Lorentz rotations accompanied with the 
 rotations in $(D-p-1)$-dim.
subspace normal to the  plane tangent to $\Sigma_{p+1}$ 
at a point $P(\mathbf{x(\xi)})$. That implies 
splitting of the local frame  into the two 
subsets: $\mathbf{n}_{A}=(\mathbf{n}_{i}, \mathbf{n}_{a})$, 
where  $\mathbf{n}_{i}, \ (i,k =0,1,...,p)$ are tangent 
 and $\mathbf{n}_{a} \ (a,b=p+1,p+2,..., D-p-1)$ are  
normal to $\Sigma_{p+1}$ at  $P(\mathbf{x(\xi)})$. 
As a result, the Cartan form  (\ref{wform}) for  the global 
Lorentz group $SO(1,D-1)$ is presented in the block form 
\begin{eqnarray}\label{spl}
\omega_{A}{}^{B}(d)
=\left( \begin{array}{cc}
                       A_{i}{}^{k}(d)& W_{ i}{}^{b}(d) \\
                         W_{ a}{}^{ k}(d) & B_{a}{}^{b}(d)
                              \end{array} \right),
\end{eqnarray}
where $(p+1)(D-p-1)$ remaining N-G bosons 
are encoded in  the bi-fundamental form $W_{ i}{}^{b}(d)$ 
playing the role of  $\omega_{G}^{i}$ in (\ref{lshi}). 
The diagonal submatrices  $A_{\mu i}{}^{k}d\xi^{\mu}$ 
and  $B_{\mu a}{}^{b}d\xi^{\mu}$ in (\ref{spl}) describe the gauge 
formes in the fundamental reps. of $SO(1,p)$
and $SO(D-p-1)$ subgroups corresponding to $\theta_{G}^{\alpha}$ 
in (\ref{lshi}), respectively. Then the integrability 
conditions (\ref{integr}) take the form 
\begin{eqnarray}
F_{\mu\nu i}{}^{k}= -(W_{[\mu} W_{\nu]})_{i}{}^{k},
\label{cF} \\
H_{\mu\nu a}{}^{b}= -(W_{[\mu} W_{\nu]})_{a}{}^{b},
\label{cH} \\
(D_{[\mu} W_{\nu]})_{i}{}^{a}=0 \label{ccd}
\end{eqnarray}
which yield the Gauss-Ricci-Codazzi (G-R-C) equations reformulated 
in terms of the massless vector multiplet  $W_{\mu  i}{}^{a}$ 
and the gauge strenghts $\hat{F}_{\mu\nu}$, $\hat{H}_{\mu\nu}$
in the  curved space $\Sigma_{p+1}$ with the induced metric   
 $g_{\mu\nu}(\xi)=\partial_{\mu}\mathbf{x}\partial_{\nu}\mathbf{x}$. 
Invariance of $\Sigma_{p+1}$ under diffeomorphisms  
requires extension of the gauge-covariant 
 derivative $(D_{\mu} W_{\nu})_{i}{}^{a}$ 
by  the Levi-Chivita connection $\Gamma_{\mu\nu}^{\rho}=\Gamma_{\nu\mu}^{\rho}$ 
\begin{eqnarray}\label{grcd}
(D_{\mu} W_{\nu})_{i}{}^{a} \ \  \rightarrow \ \
(\hat{\nabla}_{\mu}W_{\nu})_{i}{}^{a}= \partial_{\mu}W_{\nu i}{}^{a}+ A_{\mu i}{}^{k} W_{\nu k}{}^{a} + 
B_{\mu}{}^{a}{}_{b} W_{\nu i}{}^{b} - \Gamma_{\mu\nu}^{\rho}W_{\rho i}{}^{a}.
\end{eqnarray}
This covariantization does not change Eqs. (\ref{cF}-\ref{ccd}), 
but adds the Riemann-Cristoffel tensor 
\begin{eqnarray}\label{BI}
[\hat{\nabla}_{\mu} , \, \hat{\nabla}_{\nu}] 
= \hat{F}_{\mu\nu} + \hat{H}_{\mu\nu} +\hat{R}_{\mu\nu} 
\end{eqnarray}
in the commutator of the derivatives. 
The translational N-G fields induced by brane invariance
under the Poincare group $ISO(1,D-1)$ are represented 
 by the projections of $d\mathbf{x}(\xi)$
\begin{eqnarray}\label{pulb} 
\omega^{A}(d)=d\mathbf{x}(\xi)\mathbf{n}^{A}(\xi) 
\end{eqnarray}
creating the form $\omega^{A}(d)$ (\ref{mfra}) of the 
 $D$-hedron.
Encoding of the translational N-G modes $\mathbf{x}(\xi)$ 
by the forms $\omega^{A}(d)$ and $\mathbf{n}^{A}(\xi)$ is provided 
 by the integrability conditions for PDEs  (\ref{pulb}) 
\begin{eqnarray}\label{intrl}
d\wedge\omega_{A}+ \omega_{A}{}^{B}\wedge \omega_{B}=0. 
\end{eqnarray}
The orthogonality conditions $\mathbf{n}_{a}(\xi)d\mathbf{x}(\xi)=0$ 
result in the invariant constraints 
\begin{eqnarray}\label{trl}  
\omega^{a}(d)=0 \ \ \ \rightarrow \ \ \  d\mathbf{x}=\omega^{i}(d)\mathbf{n}_{i}(\xi)
\end{eqnarray}
which show  that the  quadratic element $ds^{2}=d\mathbf{x}^2$
 and $g_{\mu\nu}$ of $\Sigma_{p+1}$ are presented in the form
\begin{eqnarray}\label{metr}
ds^{2}=\omega_{i}\omega^{i}
=\omega^{i}_{\mu}\omega_{i\nu}d\xi^{\mu}d\xi^{\nu}\equiv g_{\mu\nu}(\xi)d\xi^{\mu}d\xi^{\nu}. 
\end{eqnarray}
Substitution of the condition $\omega^{a}=0$ into 
the M-C eqs. (\ref{intrl}) results in the conditions 
\begin{eqnarray}
W_{\mu i}{}^{a}= -l_{\mu\nu}{}^{a}\omega^{\nu}_{i},\ \ \ \ \ \ \ \ \ \ \ \ 
 l_{\mu\nu}{}^{a}:=\mathbf{n}^{a}\partial_{\mu\nu}\mathbf{x}, \ \ \ \ \ \ \ \ \ \ \ \ 
 \label{2frm} \\
 A_{\mu}^{ik}= \omega^{i}_{\rho}\Gamma_{\mu\lambda}^{\rho}\omega^{\lambda k} +
\omega^{i}_{\lambda}\partial_{\mu}\omega^{\lambda k},   \ \ \ \ \ \ 
F_{\mu\nu}{}^{i}{}_{k}= 
\omega^{i}_{\gamma} R_{\mu\nu}{}^{\gamma}{}_{\lambda}\omega^{\lambda}_{k},
\label{gtA'} 
\end{eqnarray}
where $l_{\mu\nu}{}^{a}$ is the second fundamental form of $\Sigma_{p+1}$.
 Eqs. (\ref{gtA'}) show that the metric connection is equivalent to the
 gauge field $A_{\mu}^{ik}$. Then invariance under $SO(1,p)$ is 
 equivalent to that under diffeomorphisms of $\Sigma_{p+1}$.
As a result,  Eqs. (\ref{cF}-\ref{ccd}) are transformed into 
\begin{eqnarray}
R_{\mu\nu}{}^{\gamma}{}_{\lambda}=l_{[\mu}{}^{\gamma a} l_{\nu]\lambda a},
\label{cRl} \\
H_{\mu\nu }{}^{ab}= l_{[\mu}{}^{\gamma a} l_{\nu]\gamma}{}^{b},
\label{cH2} \\ 
\nabla_{[\mu}^{\perp}l_{\nu]\rho a}=0, \ \ \ \ \ \
\label{ccd'}
\end{eqnarray}
where the general and  $SO(D-p-1)$ covariant  
derivative  $\nabla_{\mu}^{\perp}$ is defined as 
 \begin{eqnarray}\label{cdl}
\nabla_{\mu}^{\perp}l_{\nu\rho}{}^{a}:= \partial_{\mu}l_{\nu\rho}{}^{a}
- \Gamma_{\mu\nu}^{\lambda} l_{\lambda\rho}{}^{a} 
-\Gamma_{\mu\rho}^{\lambda} l_{\nu\lambda}{}^{a} + B_{\mu}^{ab}l_{\nu\rho b}.
 \end{eqnarray} 
 The commutator of two covariant  derivatives $\nabla^{\perp}{}_{\mu}$ (\ref{cdl})
 yields the Bianchi identitities
  \begin{eqnarray}\label{BIl}
[\nabla^{\perp}_{\gamma} , \, \nabla^{\perp}_{\nu}] l^{\mu\rho a}
=R_{\gamma\nu}{}^{\mu}{}_{\lambda} l^{\lambda\rho a}  
+ R_{\gamma\nu}{}^{\rho}{}_{\lambda} l^{\mu\lambda a} 
+H_{\gamma\nu}{}^{a}{}_{b} l^{\mu\rho b}.  
\end{eqnarray}

Thus, we show that the rotational and translational N-G fields 
are encoded by $l_{\nu\rho}{}^{a}$ and $\omega_{\mu}^{i}$. 
 The corresponding $SO(d-p-1)$ gauge invariant 
 phenomenological action of the Dirac $p$-brane in the long-wave 
 approximation, selected by  Eqs. (\ref{cRl}-\ref{ccd'})
 has the form \cite{Zbran}
\begin{eqnarray}\label{actn1}
S_{Dir.}= \gamma\int d^{p+1}\xi\sqrt{|g|}\, 
 \{
- \frac{1}{4}Sp(H_{\mu\nu}H^{\nu\mu}) 
+ \frac{1}{2}\nabla_{\mu}^{\perp}l_{\nu\rho a}\nabla^{\perp \{\mu}l^{\nu\}\rho a}
-\nabla_{\mu}^{\perp}l^{\mu}_{\rho a}\nabla_{\nu}^{\perp}l^{\nu\rho a} 
+ V_{Dir.}
\},\\
V_{Dir.}= - \frac{1}{2} Sp(l_{a}l_{b}) Sp(l^{a}l^{b})
+ Sp(l_{a}l_{b}l^{a}l^{b}) - Sp(l_{a}l^{a}l_{b}l^{b}) + c. 
\ \ \ \ \ \ \ \ \ \ \
\nonumber 
\end{eqnarray}
The action describes the interacting gauge $B_{\mu}^{ab}$ 
and tensor $l^{a}_{\mu\nu}$ 
fields  in the background $g_{\mu\nu}$. Under construction of 
the potential term $V_{Dir.}$ Eqs. (\ref{cRl}-\ref{ccd'}) have 
been taken into account together with the minimality condition 
$Spl^{a}\equiv g^{\mu\nu}l^{a}_{\mu\nu}=0$ invariant under all 
the symmetries of $S_{Dir.}$. 
The minimality conditions may be qualified as an inverse Higgs
condition \cite{IO} similarly to the conditions 
$\omega^{a}(d)=0$ (\ref{trl}) fixing the vacuum manifold for branes. 

3.  The Euler-Lagrange EOM  for  $S_{Dirac}$ (\ref{actn1}) 
are equivalently represented in the form of the second-order PDEs  
 \begin{eqnarray}
\nabla^{\perp}_{\nu} {\cal H}^{\nu\mu}_{ab}= 
 \frac{1}{2}l_{\nu\rho[a}\nabla^{\perp[\mu} l^{\nu]\rho}{}_{b ]},
  \ \ \ \ \ \ \ {\cal H}_{\mu\nu}^{ab}:= H_{\mu\nu}^{ab} 
- l_{[\mu}{}^{\gamma a} l_{\nu]\gamma}{}^{b},
\label{maxH2*} \\
\nabla^{\perp}_{\mu}\nabla^{\perp[\mu}l^{\nu]\rho a}=0,  \ \ \ \ \ 
\ \ \ \ \ \ \ \ \ \ \ \ \ \ \ 
\label{eqgW2*}
\end{eqnarray}
where  ${\cal H}_{\mu\nu}^{ab}$ is the shifted 
strenght $H_{\mu\nu}^{ab}$ which presents
Eqs. (\ref{cH2}) and   (\ref{ccd'}) as
\begin{eqnarray}\label{Hlsol} 
{\cal H}_{\mu\nu}^{ab}(\tau,\sigma^r)=0, \ \ \ \ \  
\nabla_{[\mu}^{\perp}l_{\nu]\rho a}(\tau,\sigma^r)=0.
\end{eqnarray} 
 Noting that  Eqs. (\ref{Hlsol}) are  the first order PDEs
 we will prove that they can be interpreted as the Cauchy initial 
data for PDEs (\ref{maxH2*}) and (\ref{eqgW2*}). 
For this purpose let us consider Eqs. (\ref{Hlsol}) 
as some constraints chosen at the time  $\tau=0$:
 \begin{eqnarray}\label{cauchy}
{\cal H}^{\nu\mu}_{ab}(0,\sigma^r)=0, \ \ \ \
\nabla^{\perp[\mu}l^{\nu]\rho a}(0,\sigma^r)=0   
\end{eqnarray}
and analyze their time evolution prescribed by 
EOM (\ref{maxH2*}), (\ref{eqgW2*}) considering the expansion
 \begin{eqnarray}
 {\cal H}^{\tau r}_{ab}(\delta\tau,\sigma^r)= {\cal H}^{\tau r}_{ab}(0,\sigma^r ) 
 + \partial_{\tau} {\cal H}^{\tau r}_{ab}|_{\tau=0}\delta\tau +... ,
 \nonumber 
\\
 \nabla^{\perp[\tau}l^{r]\rho a}(\delta\tau,\sigma^r )=\nabla^{\perp[\tau}l^{r]\rho a}(0,\sigma^r )
 +\partial_{\tau} \nabla^{\perp[\tau}l^{r]\rho a}|_{\tau=0}\delta\tau +... 
 \label{varid}
 \end{eqnarray}
Using EOM (\ref{maxH2*}), (\ref{eqgW2*}) and the initial data 
constraints (\ref{cauchy}) we transform  Eqs.(\ref{varid}) as follows
 \begin{eqnarray}
 {\cal H}^{\tau r}_{ab}(\delta\tau,\sigma^r)= 
 -\nabla^{\perp}_{r'} {\cal H}^{r' r}_{ab}|_{\tau=0}\delta\tau+... ,
 \nonumber 
\\
 \nabla^{\perp[\tau}l^{r]\rho a}(\delta\tau,\sigma^r )=
 -\nabla^{\perp}_{r'}\nabla^{\perp[r'}l^{r]\rho a}|_{\tau=0}\delta\tau +... 
 \label{varid'}
 \end{eqnarray}
 Observing that the  space covariant derivatives 
 of (\ref{cauchy}) are equal to zero 
\begin{eqnarray}\label{cauchy'} 
\nabla^{\perp}_{r'}{\cal H}^{\nu\mu}_{ab}(0,\sigma^r)=
\nabla^{\perp}_{r'}\nabla^{\perp[\mu}l^{\nu]\rho a}(0,\sigma^r)=0,
 \ \ \ \ (r'=1,2,...,p-1) 
\end{eqnarray} 
we find that Eqs. (\ref{Hlsol}) are  conserved in time in view of the dynamics 
given by  (\ref{maxH2*}), (\ref{eqgW2*}), the first pair of the extended 
Maxwell equations and the Ricci identities 
\begin{eqnarray}\label{Hlsol'} 
{\cal H}_{\mu\nu}^{ab}(\delta\tau,\sigma^r) ={\cal H}_{\mu\nu}^{ab}(0,\sigma^r ), \ \ \ \ \  
\nabla_{[\mu}^{\perp}l_{\nu]\rho a}(\delta\tau,\sigma^r)=\nabla_{[\mu}^{\perp}l_{\nu]\rho a}(0,\sigma^r). 
\end{eqnarray} 
Using the Cauchy-Kowalevskaya theorem of local 
existence and iniqueness we conclude that Eqs. (\ref{Hlsol}) present the 
generic solution of EOM (\ref{maxH2*}-\ref{eqgW2*})
corresponding to the initial data (\ref{cauchy}). 
The solution selects  the closed  sector of evolution states
 lying on the surface of the constraints (\ref{Hlsol}) under an arbitrary 
 but fixed metric $g_{\mu\nu}$. In its turn, the metric dynamics 
 is defined  by the Gauss condition (\ref{cRl}) treated 
 as the second-order PDEs for $g_{\mu\nu}$ with the 
 given $l^{a}_{\mu\nu}$. In its 
 turn the dynamics of the N-G 
 multiplet $l^{a}_{\mu\nu}$ is derived using the standard 
 variational principle for the action (\ref{actn1}).
For the Nambu-Goto string (p=1) in 3-dimensional Minkowski
space the metric condition (\ref{cRl}) is the Lioville equation
proving that the string world-sheet is 2-dim. Einstein space with 
$R_{\mu\nu}=\kappa g_{\mu\nu}$ as we will show in the next section. 

So, we prove that the action (\ref{actn1}) desribes the 
Dirac $p$-brane dynamics encoded in the quartic potential 
$V_{Dir.}$ selected by the G-R-C  constraints (\ref{cRl}-\ref{ccd'}) 
with $Spl^{a}=0$. 
 Note that a flat brane with $g_{\mu\nu}=\eta_{\mu\nu}$ 
corresponds to the evident  solution $l_{\mu\nu}^{a}=0$ and $V_{Dir.}=const$.   

4. Here we illustrate the work  of the above-discussed approach
for  a simple case of string ($p=1$) embedded in 3-dim. Minkowski space 
 studied in \cite{Zgau24} (see also \cite{BN}. 
Using the gauge $l_{\tau\tau}=l_{\sigma\sigma}=0$ and the minimality condition 
$Sp\ l$=0 (where $l_{\mu\nu}\equiv l_{\mu\nu 2}$) we find
\begin{eqnarray}\label{lig}
l_{\mu\nu}
=\left( \begin{array}{cc}
                      0& M\\
                     M & 0
\end{array} \right),
\ \ \ \ \   Sp \ l=0  \ \   \rightarrow \ \ 
g_{\mu\nu}
=\left( \begin{array}{cc}
               g_{\tau}& 0\\
                     0 & g_{\sigma} 
\end{array} \right), \ \ \ \
M\ne 0.
\end{eqnarray}
In any 2-dim. space the Riemann tensor has only one non-zero 
component $R_{\tau\sigma\tau\sigma}$. Then the Gauss equation (\ref{cRl}) 
 are reduced to $R_{\tau\sigma\tau\sigma}=(l_{\tau\sigma})^2\equiv M^2$,
 and we obtain 
\begin{eqnarray}\label{ric}
R_{\mu\nu}
= M^{2}\left( \begin{array}{cc}
                   1/g_{\sigma} & 0\\
                    0 & 1/g_{\tau} 
\end{array} \right),
\ \ \ \ \  R=2\frac{M^{2}}{g_{\tau}g_{\sigma}}=-2\frac{detl_{\mu\nu}}{detg_{\mu\nu}}            
\end{eqnarray}
for the Ricci tensor $R_{\nu\lambda}=g^{\mu\gamma}R_{\mu\nu\gamma\lambda}$ 
 and the scalar curvature $R=g^{\mu\nu}R_{\mu\nu}$ of the worldsheet.
From Eqs. (\ref{lig}),  (\ref{ric}) we obtain the 
equation for 2-dim. Einstein space geometry 
\begin{eqnarray}\label{HE}
  R_{\mu\nu}-\frac{1}{2}g_{\mu\nu}R=0
\end{eqnarray}
defining dynamics  of the worldsheet metric $g_{\mu\nu}$ 
in the reformulated Nambu-Dirac action (\ref{actn1}). Eq. 
 (\ref{cH2}) is identically satisfied since $B^{ab}_{\mu}\equiv 0$.  
 Finally, Eqs. (\ref{ccd'}) take the following form 
\begin{eqnarray}
\partial_{[\mu}l_{\nu]\rho}- \Gamma_{\rho[\mu}^{\lambda}l_{\nu]\lambda}=0, \ \ \ \ \ \ \ \ \ \ \ \ 
\ \ \ \ \ \ \ \ \ \ \ \ \ \ \ \ \ 
\label{cc2}\\
\Gamma_{\tau\beta}^{\alpha}
=\frac{1}{2}\left(\begin{array}{cc}
                   \partial_{\tau}g_{\tau}/g_{\tau} & \partial_{\sigma}g_{\tau}/g_{\tau}\\
              -\partial_{\sigma}g_{\tau}/g_{\sigma} & \partial_{\tau}g_{\sigma}/g_{\sigma},              
  \end{array} \right),  \ \ \  \  \label{crtf}
\Gamma_{\sigma\beta}^{\alpha}
=\frac{1}{2}\left(\begin{array}{cc}
                   \partial_{\sigma}g_{\tau}/g_{\tau} & -\partial_{\tau}g_{\sigma}/g_{\tau}\\
              \partial_{\tau}g_{\sigma}/g_{\sigma} & \partial_{\sigma}g_{\sigma}/g_{\sigma}             
\end{array} \right),             
\end{eqnarray}
where $\alpha$ and $\beta$ number the rows and columns. 
Insertion of (\ref{crtf}) into (\ref{cc2}) reduces it to  
\begin{eqnarray}\label{cc2'}
\partial_{\tau}M/M - \frac{1}{2}( \partial_{\tau}g_{\tau}/g_{\tau} - \partial_{\tau}g_{\sigma}/g_{\sigma})=0, \ \ \
\partial_{\sigma}M/M + \frac{1}{2}( \partial_{\sigma}g_{\tau}/g_{\tau} - \partial_{\sigma}g_{\sigma}/g_{\sigma})=0. \ \ \
\end{eqnarray} 
The integrability condition of (\ref{cc2'})  
gives  $g_{\sigma}=-e^{2[-\varphi(\tau)+\chi(\sigma)]}g_{\tau}$ 
and we get the general solution of (\ref{cc2'}) 
$M=\pm {\tilde k}e^{\varphi(\tau)+\chi(\sigma)}, \,  ({\tilde k}=constant)$. 
The presence of arbitrary  "integration constants" $\varphi(\tau)$ 
and $\chi(\sigma)$ in the solution is a consequence of the 
 gauge symmetry of Eqs. (\ref{cc2'})
\begin{eqnarray}\label{weyl}
M'=e^{\alpha(\tau) + \beta(\sigma)}M, \ \ \ \ \ \ \ \ 
(g_{\sigma}/g_{\tau})'= 
e^{-2[(\alpha(\tau) + \delta(\tau))- (\beta(\sigma) + \gamma(\sigma))]}(g_{\sigma}/g_{\tau}),
\end{eqnarray}
where $\alpha, \beta, \gamma, \delta$ are arbitrary functions, 
as folows from Eqs. (\ref{cc2'}) represented in the form: 
$\partial_{\tau}ln(M^2|g_{\sigma}/g_{\tau}|)=0, \ \ \
\partial_{\sigma}ln(M^2|g_{\tau}/g_{\sigma}|)=0$.
So, one can choose the conformal gauge
\begin{eqnarray}\label{confo}
g_{\sigma}=-g_{\tau} \equiv -E(\tau, \sigma), \ \ \ \  M^2=k,  \ \ \ \   R_{\tau\sigma\tau\sigma}=k
\end{eqnarray}
So, the rotational N-G field $M$ is a solution of 
Eq. $\Box M=0$ for a massless
 scalar field in the 2-dim. conformal flat space-time. 
 To solve the H-E Eq. (\ref{HE})
we use the definition 
\begin{eqnarray}\label{HE1}
R_{\tau\sigma\tau\sigma}=ER_{\tau\sigma}{}^{\tau}{}_{\sigma}=E(\partial_{[\tau} \Gamma_{\sigma]\sigma}^{\tau}
+\Gamma_{\rho[\tau}^{\tau}\Gamma_{\sigma]\sigma}^{\rho})
=\frac{1}{2E}[E\Box E -(\partial_{\tau}E)^2 + (\partial_{\sigma}E)^2]
\end{eqnarray}
for the Riemann tensor and representation (\ref{crtf}) referred to 
the conformal gauge. Making the  change
$E:= e^{2\psi}$ and using Eq.
(\ref{confo}) we transform  
 Eq. (\ref{HE}) into the Lioville one
\begin{eqnarray}\label{HELi}
(\partial_{\tau}^2-\partial_{\sigma}^2)\psi=ke^{-2\psi}
\end{eqnarray}
 earlier proved to describe the relativistic string ($p=1, D=3$) in 
 the geometrical approach \cite{BNCh}. 
 This shows the equivalence of the gauge approach 
to the standard approach resulting in 2-dim. conformal 
invariant EOM (\ref{HELi}) for the string metrics on the classical level.

\medskip  

{\bf Summary}: 

 The Namby-Goldstone fields of the spontaneously broken
  internal symmetries were described as the 
 effective massless Yang-Mills-Cartan multiplets.
The interpretation of the Dirac $p$-brane \cite{Zbran} 
embedded into $\mathbf{R}^{1,D-1}$ 
as a dynamical system with the $ISO(1,D-1)$
symmetry spontaneously broken to $ISO(1,p)\times SO(D-p-1)$
was established. The brane metric dynamics was shown to be
 described by the generalized eqs. of 
the {\it Gauss Theorema Egregium}. 
 For the case of string 
embedded into $\mathbf{R}^{1,2}$ the geometry of its worldsheet turned out 
 to coincide with the  geometry of 2-dim. Einstein space.  
Having in mind a connection between the {\it quantum} conformal 
invariance in string theory,  vanishing  of its beta function and the vacuum 
H-E eqs. \cite{Lov}, \cite{CFMP} it seems  interesting to find its generalization to branes.

\bigskip 

\noindent{\bf Acknowledgments}

I am grateful to NORDITA and Physics Department 
of Stockholm University for kind hospitality and support.
These results were reported at the  Workshop SQS-2015
 "Supersymmetries and Quantum Symmtries" in memory of V.I. Ogievetsky.

\end{document}